\documentclass[twocolumn,showpacs,preprintnumbers,amsmath,amssymb]{revtex4}

\usepackage{graphicx}
\usepackage{dcolumn}
\usepackage{bm}
\begin{document}
\preprint{Nuclear Astrophysics}
\title{Charged Lepton Production from Iron Induced by Atmospheric Neutrinos}
\author{M. Sajjad Athar, S.Ahmad and S.K.Singh}
\email{pht13sks@rediffmail.com}
\affiliation{Department of Physics, Aligarh Muslim University, Aligarh-202 002, India.}
\date{\today}
 
\begin{abstract}
 The charged current lepton production induced by neutrinos in $^{56}Fe$ nuclei has been studied. The calculations have been done for the quasielastic as well as the inelastic reactions assuming $\Delta$ dominance and take into account the effect of Pauli blocking, Fermi motion and the renormalization of weak transition strengths in the nuclear medium. The quasielastic production cross section for lepton production are found to be strongly reduced due to nuclear effects while there is about $10\%$ reduction in the inelastic cross sections in the absence of the final state interactions of the pions. The numerical results for the momentum and angular distributions of the leptons averaged over the various atmospheric neutrino spectra at the Soudan and Gransasso sites have been presented. The effect of nuclear model dependence and the atmospheric flux dependence on the relative yield of ${\mu}$ to e has been studied and discussed.
\end{abstract}

\maketitle 

\section{Introduction}                 
The study of neutrino physics with atmospheric neutrinos has a long history with first observations of muons produced by atmospheric muon neutrinos in deep underground laboratories of KGF in India and ERPM in South Africa\cite{achar}. The indications of some deficit in the atmospheric neutrino flux was known to exist from the early days of these experiments but the evidence was no more than suggestive due to low statistics of the experimental data and anticipated uncertainties in the flux calculations\cite{crouch}. The clear evidence of a deficit in the atmospheric muon neutrino flux was confirmed later when data with better statistics were obtained at IMB\cite{heines}, Kamiokande\cite{hirata} and Soudan\cite{allison} experiments. The most likely cause of this deficit is believed to be the phenomena of neutrino oscillations\cite{pontecarvo} in which the neutrinos produced with muon flavor after passing a certain distance through the atmosphere, manifest themselves as a different flavor. The implication of this phenomena of neutrino oscillation is that neutrinos possess a nonzero mass pointing towards physics beyond the standard model of particle physics. The evidence for neutrino oscillations and a nonzero mass for the neutrinos has also been obtained in the observations made with solar\cite{solar} and reactor (anti)neutrinos\cite{reactor}.

It is well known that, in a two flavor oscillation scenario involving muon neutrino, the probability for a muon neutrino with energy $E_\nu$ to remain a muon neutrino after propagating a distance $L$ before reaching the detector is given by\cite{pontecarvo}. 
\begin{equation}
P_{\mu\mu}=1-sin^2 2\theta sin^2\left(\frac{1.27\Delta{m^2}(eV^2)L(km)}{E_\nu(GeV)}\right) 
\end{equation}
where $\Delta{m^2}=m^2_1-m^2_2$ is the difference of the squared masses of the two flavor mass eigenstates and $\theta$ is the mixing angle between two states. The oscillation parameters $\Delta{m^2}$ and the mixing angle $\theta$ are determined by various observations made in atmospheric neutrino experiments. These include the flavor ratios of muon and electron flavors, angular and $\frac{L}{E}$ distributions of muons and electrons produced by atmospheric neutrinos. The first claims of seeing neutrino oscillation in atmospheric neutrinos came from the measurements of ratio of ratios $R_\nu$ defined as $\frac{(\mu/e)_{data}}{(\mu/e)_{MC}}$ from the observations of fully contained (FC) events\cite{heines}-\cite{hirata}, but there are now data available from the angular and $\frac{L}{E}$ distribution of the atmospheric neutrino induced muon and electron events from SK\cite{sk}, MACRO\cite{macro} and Soudan\cite{soudan} experiments which confirm the phenomena of neutrino oscillations. These experiments are consistent with a value of $\Delta{m^2}\approx 3.2\times 10^{-3}eV^2$ and $sin^2 2\theta\approx 1$. The analysis of these data assuming a three flavor neutrino oscillation phenomenology have also been done by many authors\cite{threefla}.

The major sources of uncertainty in the theoretical prediction of the charged leptons of muon and electron flavor produced by the atmospheric neutrinos come from the uncertainties in the calculation of atmospheric neutrino fluxes and neutrino nuclear cross sections. The atmospheric neutrino fluxes at various experimental sites of Kamioka, Soudan and Gransasso have been extensively discussed in literature by many authors\cite{battistoni}-\cite{plyaskin}. The neutrino nuclear cross sections have also been calculated for various nuclei by many authors using different nuclear models\cite{donnelly}-\cite{singh1}. The aim of the present paper is to study the neutrino nuclear cross section in iron nuclei which are relevant for the atmospheric neutrino experiments performed at Soudan\cite{soudan}, FREJUS\cite{frejus} and NUSEX\cite{nusex} and planned in future with MINOS\cite{minos}, MONOLITH\cite{monolith} and INO\cite{ino} detectors. The uncertainty in the nuclear production cross section of leptons from iron nuclei by the atmospheric neutrinos are discussed. For our nuclear model, we also discuss the uncertainty due to use of different neutrino fluxes for the sites of Soudan and Gransasso which are relevant to MINOS, MONOLITH and INO detectors\cite{minos}-\cite{ino}.

The momentum and angular distribution of muons and electrons relevant to fully contained events produced by atmospheric neutrinos in iron nuclei are calculated. These leptons of muon and electron flavor characterized by track and shower events include the leptons produced by quasielastic process as well as the inelastic processes induced by charged current interactions. The calculations are done in a model which takes into account nuclear effects like Pauli Blocking, Fermi motion effects and the effect of renormalization of the weak transition strengths in nuclear medium in local density approximation. The model has been successfully applied to describe various electromagnetic and weak processes like photon absorption, electron scattering, muon capture and low energy neutrino reactions in nuclei\cite{singh1},\cite{gil}-\cite{singh4}. The model can be easily applied to calculate the zenith angle dependence and the $\frac{L}{E}$ distribution for stopping and thorough going muon production from iron nuclei which is currently under progress.

The plan of the paper is as follows. In section-II we describe the neutrino(antineutrino) quasielastic inclusive production of leptons $(e^-, \mu^-, e^+, \mu^+)$ from iron nuclei for various neutrino energies. In section-III we describe the energy dependence of the inelastic production of leptons through $\Delta$-dominance model and highlight the nuclear effects relevant to the energy of fully contained events. In section-IV, we use the atmospheric neutrino flux at Soudan and Gransasso sites as determined by various authors and discuss the flux averaged momentum and angular dependence of leptons corresponding to different flux calculations available at these two sites.
\section{Quasielastic Production of Leptons}
Quasielastic inclusive production of leptons in nuclei induced by neutrinos has been studied by many authors where nuclear effects have been calculated. Most of these calculations have been done either for $^{16}{O}$ relevant to IMB and Kamioka experiments\cite{heines}-\cite{hirata}or for $^{12}{C}$ relevant to LSND and KARMEN experiment\cite{lsnd1}. These calculations generally use direct summation method (over many nuclear excited states)\cite{donnelly}, closure approximation\cite{goulard}, Fermi gas model\cite{smith}-\cite{smith1}, relativistic mean field approximation\cite{kim}, continuum random phase approximation (CRPA)\cite{engel} and local density approximation\cite{singh2,singh1}. The calculations for $^{56}{Fe}$ nucleus have been reported by Bugaev et al.\cite{donnelly} in a shell model and in Fermi gas model by Gallagher\cite{gallagher} and Berger et al.\cite{berger}. In this section we briefly describe the formalism and results of our calculations done for quasielastic inclusive production of leptons for iron nuclei.
\subsection{Formalism}
In local density approximation the neutrino nucleus cross section $\sigma(E_\nu)$ for a neutrino of energy $E_\nu$ scattering from a nucleus $A(Z,N)$, is given by
\begin{equation}
\sigma^A(E_\nu)=2\int d\vec{r}\frac{d\vec{p}}{(2\pi)^3}n_n(\vec{p}, \vec{r})\sigma^N(E_\nu)
\end{equation}
where $n_n(\vec{p}, \vec{r})$ is the local occupation number of the initial nucleon of momentum $\vec{p}$ (localized at position $\vec{r}$ in the nucleus) and $\sigma^N(E_\nu)$ is cross section for the scattering of neutrino of energy $E_\nu$ from a free nucleon given by the expression
\begin{eqnarray}
\sigma^N(E_\nu)&=&\int\frac{d^3k^\prime}{(2\pi)^3}\frac{m_\nu}{E_\nu} \frac{m_e}{E_e} \frac{M_n}{E_n} \frac{M_p}{E_p}\nonumber\\
&\times&\bar{\sum}\sum|T|^2\delta(E_\nu-E_l+E_n-E_p) 
\end{eqnarray}
where $T$ is the matrix element for the basic process 
\begin{equation}
\nu_l(k)+n(p)\rightarrow l^-(k^\prime)+p(p^\prime),~~l=e,\mu
\end{equation}
written as
\begin{equation}
T = \frac{G_F}{\sqrt{2}}\cos{\theta_c} ~\bar{u}(k^\prime)\gamma_\mu(1-\gamma_5)u(k) ~{(J^{\mu})}^{CC}
\end{equation}
${(J^\mu)}^{CC}$ is the charged current(CC) matrix element of the hadronic current defined as
\begin{eqnarray}
{(J^\mu)}^{CC}=\bar{u}(p^\prime)\left[{F_{1}^V}(q^2)\gamma^\mu+ {F_{2}^V}(q^2)i{\sigma^{\mu\nu}}{\frac{q_\nu}{2M}}\right.\\ \nonumber
\left. + {F_{A}^V}(q^2)\gamma^\mu\gamma^5\right]u(p)
\end{eqnarray}
$q^2(q=k-k^\prime)$ is the four momentum transfer square. The form factors $F_1^V(q^2)$, $F_2^V(q^2)$ and $F_A^V(q^2)$ are isovector electroweak form factors written as
\[F_1^V(q^2)=F_1^p(q^2)-F_1^n(q^2), 
~F_2^V(q^2)=F_2^p(q^2)-F_2^n(q^2),\]
\[F_A^V(q^2)=F_A(q^2)\]
where
\begin{eqnarray*}
 F^{p,n}_{1}(q^2)&=&\frac{1}{(1-\frac{q^2}{4M^2})}\left[{G^{p,n}_E(q^2)-\frac{q^2}{4M^2}G^{p,n}_M(q^2)}\right]\\
 F^{p,n}_2(q^2)&=&\frac{1}{(1-\frac{q^2}{4M^2})}[{G^{p,n}_M(q^2)-G^{p,n}_E(q^2)}]
\end{eqnarray*}
\begin{equation}
G^p_E(q^2)=\left(1-\frac{q^2}{M^2_v}\right)^{-2}
\end{equation}
\[G^p_M(q^2)=(1+\mu_p)G^p_E(q^2),~G^n_M(q^2)=\mu_n G^p_E(q^2);  \]
\[G^n_E(q^2)=(\frac{q^2}{4M^2})\mu_n G^p_E(q^2) \xi_n;~\xi_n=\frac{1}{1- \lambda_n\frac{q^2}{4M^2}}\]
\[\mu_p=1.79, \mu_n=-1.91, M_v=0.84GeV, ~\mbox{and}  ~\lambda_n=5.6.\]
The isovector axial vector form factor ${F_A}(Q^2)$ is given by
\[ {F_A}(Q^2)=\frac{F_A(0)}{({1-\frac{q^2}{M^2_A})}^2}\]
where $M_A=1.032GeV$; $F_A(0)$=-1.261

In a nuclear process the neutrons and protons are not free and their momenta are constrained by the Pauli principle which is implemented in this model by requiring that for neutrino reactions initial nucleon momentum $p\le p_{F_n}$ and final nucleon momentum $p^\prime=(|\vec{p}+\vec{q}|)>p_{F_p}$ where $p_{F_{n,p}}=[\frac{3}{2}\pi^2\rho_{n,p}(r)]^{\frac{1}{3}}$, are the local Fermi momenta of neutrons and protons at the interaction point in the nucleus defined in terms of their respective nuclear densities $\rho_{n,p}(r)$. These constraints are incorporated while performing the integration over the initial nucleon momentum in Eqn.(2) by replacing the energy conserving $\delta$-function in Eqn.(3) by $-\frac{1}{\pi}Im U_N(q_0, \vec{q})$ where $U_N(q_0, \vec{q})$ is the Lindhard function corresponding to the particle hole(ph) excitations induced by the weak interaction process through W exchange shown in Fig.1(a). In the large mass limit of W boson i.e.$M_W\rightarrow \infty$, this Fig.1(a) is reduced to Fig.1(b) for which the imaginary part of the Lindhard function i.e. $Im U_N(q_0, \vec{q})$ is given by
\begin{figure}[h]
\includegraphics{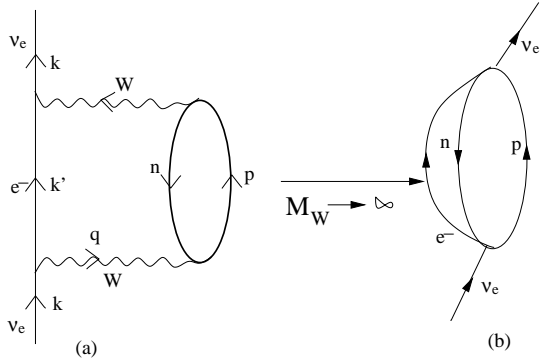}
\caption{Diagrammatic representation of the neutrino self-energy diagram corresponding to the ph-excitation leading to $\nu_e +n \rightarrow e^- + p$ process in nuclei. In the large mass limit of the $W$ boson (i.e.$M_W\rightarrow \infty$) the diagram 1(a) is reduced to 1(b) which is used to calculate $|T|^2$ in Eqn.(5).}
\end{figure}
\begin{equation}
Im U_N(q_0, \vec{q})=-\frac{1}{2\pi}\frac{M_p{M_n}}{|\vec{q}|}\left[E_{F_1}-A\right]~\mbox{with}
\end{equation}
\[q^2<0,~E_{F_2}-q_0<E_{F_1} ~\mbox{and} ~\frac{-q_0+|\vec{q}|{\sqrt{1-\frac{4{M^2}}{q^2}}}}{2}<{E_{F_1}}\] 
where $E_{F_1}$ and $E_{F_2}$ are the local Fermi energy of initial and final nucleons and 
\[A = Max\left[M_n,\hspace{2mm}E_{F_2}-q_0,\hspace{2mm}\frac{-q_0+|\vec{q}|\sqrt{1-\frac{4{M^2}}{q^2}}}{2}\right]\]
The expression for the neutrino nuclear cross section $\sigma^A(E_\nu)$, is then given by:
\begin{eqnarray}
\sigma^A(E_\nu)=-\frac{4}{\pi}\int^{r_{max}}_{r_{min}} r^2 dr \int^{{p_l}^{max}}_{{p_l}^{min}}{p_l}^2dp_l 
\int_{-1}^1d(cos\theta)\nonumber\\
\times\frac{1}{E_\nu E_l} {\bar{\sum}}\sum|T|^2 Im{U_N}[E_\nu-E_l,~\vec{q}].
\end{eqnarray}
Moreover in the nucleus, the $Q$ value of the nuclear reaction and the Coulomb distortion of the final lepton in the electromagnetic field of the final nucleus should be taken into account. This is done by modifying the energy conserving $\delta$-function $\delta(E_\nu-E_l+E_n-E_p)$ in Eqn.(3) to $\delta(E_\nu-Q-(E_l+V_c(r))+E_n-E_p)$ where $V_c(r)$ is the Coulomb energy of the produced lepton in the field of final nucleus and is given by
\begin{equation}
V_c(r)=ZZ^\prime\alpha4\pi(\frac{1}{r}\int_0^r\frac{\rho_p(r^\prime)}{Z}{r^\prime}^2dr^\prime + \int_r^\infty\frac{\rho_p(r^\prime)}{Z}{r^\prime}dr^\prime)
\end{equation}

This amounts to evaluation of Lindhard function in Eqn.(8) at $(q_0-(Q+V_c(r)),~\vec{q})$ instead of $(q_0, ~\vec{q})$. The implementation of this modification requires a judicious choice of $Q$ value for inclusive nuclear reactions in which many nuclear states are excited in iron. We have taken $Q$-value of 6.8MeV corresponding to the transition to lowest lying $1^+$ state in $^{56}{Co}$ for $\nu_l+^{56}{Fe}\rightarrow l^-+^{56}{Co^\star}$ reaction and $Q$-value of 4.3MeV  corresponding to the transition to lowest lying $1^+$ state in $^{56}{Mn}$ for ${\bar{\nu_l}}+^{56}{Fe}\rightarrow l^++^{56}{Mn^\star}$ reaction. 

The inclusion of $V_c(r)$ to modify energy and corresponding momentum of the charged lepton in the Coulomb field of final nucleus in our model is equivalent to the treatment of Coulomb distortion effect in modified effective momentum approximation(MEMA). This approximation has been used in other calculations of charged current neutrino reactions\cite{volpe} and electron scattering at higher energies\cite{guisti}. 

With these modifications, the final expression for the quasielastic inclusive production from iron nucleus is given by
\begin{eqnarray}
\sigma^A(E_\nu)=-\frac{4}{\pi}\int^{r_{max}}_{r_{min}} r^2 dr \int^{{p_l}^{max}}_{{p_l}^{min}}{p_l}^2dp_l 
\int_{-1}^1d(cos\theta)\nonumber\\
\times\frac{1}{E_\nu E_l} {\bar{\sum}}\sum|T|^2 Im{U_N}[q_0-(Q+V_c(r)),~\vec{q}].
\end{eqnarray}
It is well known that weak transition strengths are modified in the nuclear medium due to presence of strongly interacting nucleons. This modification of the weak transitions strength in the nuclear medium is taken into account by considering the propagation of particle-hole(ph) excitations in the medium. While propagating through the medium, the ph-excitations interact through the nucleon nucleon potential and create other particle hole and $\Delta$-h excitations as shown in Fig.2. 
\begin{figure}[h]
\includegraphics{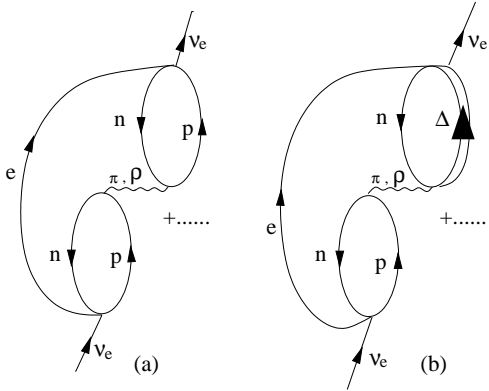}
\caption{ Many body Feynman diagrams (drawn in the limit $M_W\rightarrow \infty$) accounting for the medium polarization effects contributing to the process $\nu_e +n \rightarrow e^- + p$ transitions.}
\end{figure}
The effect of these excitations are calculated in random phase approximation which is described in Ref.\cite{singh2,singh4}. The effect of nuclear medium on the renormalization of weak strengths is treated in a non-relativistic frame work. In leading order, the non-relativistic reduction of the weak hadronic current defined in Eqn.6, $F_2(q^2)$ term gives a spin-isospin transition operator $\frac{\vec{\sigma}\times{\vec{q}}}{2M}\vec{\tau}$ which is a transverse operator while the $F_A(q^2)$ term gives a spin-isospin transition operator $\vec{\sigma}\cdot\vec{\tau}$ which has a longitudinal as well as a transverse part. This representation of the transition operators in the longitudinal and transverse parts is useful when summing the diagrams in Fig.(2) in random phase approximation(RPA) to calculate $|T|^2$. While the charge coupling remains unchanged due to nuclear medium effects, the terms proportional to $F_2^2$ are affected by the transverse part of the nucleon nucleon potential while the terms proportional to $F_A^2$ are affected by transverse as well as longitudinal parts. The effect is to replace the terms like $F_2^2$, $F_A^2$, $F_2F_A$ etc. in the following manner\cite{singh2,singh4}
\begin{eqnarray}
(F^2_{2},F_2F_A) \rightarrow (F^2_{2},F_2F_A)\frac{1}{|1-U_N{V_t}|^2}\nonumber\\
F^2_A  \rightarrow  \left[\frac{1}{3}\frac{1}{{|1-U_N{V_l}|}^2}+\frac{2}{3}\frac{1}{{|1-U_N{V_t}|}^2}\right]
\end{eqnarray}
where $V_l$ and $V_t$ are the longitudinal and transverse parts of the nucleon nucleon potential calculated with $\pi$ and $\rho$ exchanges and modulated by Landau-Migdal parameter $g^\prime$ to take into account the short range correlation effects and are given by
\begin{eqnarray}
V_l(q) = \frac{f^2}{m_\pi^2}\left[\frac{q^2}{-q^2+m_\pi^2}{\left(\frac{\Lambda_\pi^2-m_\pi^2}{\Lambda_\pi^2-q^2}\right)^2}+g^\prime\right],\nonumber\\
V_t(q) = \frac{f^2}{m_\pi^2}\left[\frac{q^2}{-q^2+m^2_\rho}{C_\rho}{\left(\frac{{\Lambda_\rho}^2-m^2_\rho}{{\Lambda_\rho}^2-q^2}\right)^2}+g^\prime\right]\end{eqnarray}
with $\Lambda_\pi=1.3 GeV$, $C_\rho=2.0$, $\Lambda_\rho=2.5GeV$, $m_\pi$ and $m_\rho$ are the pion and rho meson masses and $g^\prime$ is taken to be $0.7$\cite{mukh}. The effect of $\Delta$h excitations are taken into account by including the Lindhard function $U_{\Delta}$ for the $\Delta$h excitations and replacing $U_N$ by $U_N+U_{\Delta}$ in Eqn.(12). The complete expression for $U_N$ and $U_\Delta$ used in our calculations are taken from \cite{oset1}. The different couplings for $N$ and $\Delta$ to the nucleon are incorporated in $U_N$ and $U_{\Delta}$ and then the same interaction strengths $V_l$ and $V_t$ are used for ph and $\Delta$h excitations\cite{garcia}-\cite{salcedo}.
\subsection{Results}
We present the numerical results for the total cross section for the quasielastic processes $\nu_l({\bar{\nu_l}})+^{56}{Fe}\rightarrow l^-(l^+)+^{56}{Co^\star}(^{56}{Mn^\star})$ as a function of energy for neutrino and anti neutrino reactions on iron in the energy region relevant to the fully contained events of atmospheric neutrinos i.e. $E_\nu<3 GeV$. The cross sections have been calculated using Eqn.(11) with the nuclear density $\rho(r)$ given by a two parameter Fermi density\cite{vries}:
 $\rho(r)=\frac{\rho(0)}{1.+exp(\frac{r-c}{z})}$ with $c=3.971fm$, $z=0.5935fm$, $\rho_n(r)=\frac{(A-Z)}{A}\rho(r)$ and $\rho_p(r)=\frac{Z}{A}\rho(r)$.

In Fig.3 we show the numerical results of $\sigma(E)$ vs $E$, for all flavors of neutrinos i.e. $\nu_\mu$, $\bar{\nu}_\mu$, $\nu_e$ and $\bar{\nu}_e$. The reduction due to nuclear effects is large at lower energies but becomes small at higher energies. The energy dependence of the cross sections for muon and electron type neutrinos are similar except for the threshold effects which are seen only at low energies $(E_\nu < 500MeV)$. This reduction in $\sigma$ is due to Pauli blocking as well as due to the weak renormalization of transition strengths which have been separately shown in Fig.4(a) and Fig.4(b) for neutrinos and antineutrinos, where we also show the results in the Fermi gas model given by Llewellyn Smith\cite{smith1}. We plot in Fig.4(a) and Fig.4(b) for neutrinos and antineutrinos, the reduction factor $R=\frac{\sigma_{nuclear}(E)}{\sigma_{nucleon}(E)}$ vs E where $\sigma_{nuclear}(E)$ is the cross section per neutron(proton) for neutrino(antineutrino) reactions in the nuclear medium. The solid lines show the reduction factor R when only the Pauli suppression is taken into account through the imaginary part of the Lindhard function given in Eqn.8. This is similar to the results of Llewellyn Smith\cite{smith1} in Fermi gas model shown by dashed lines.  In this model the total cross section $\sigma$ is calculated by using the formula $\sigma=\int dq^2~R(q^2)(\frac{d\sigma}{dq^2})_{free}$ where $R(q^2)$ describes the reduction in the cross section calculated in the Fermi gas model and includes the effect of Pauli suppression only\cite{smith1}. However, in our model we get further reduction due to renormalization of weak transition strengths in the nuclear medium when the effects of Fig.2(a) and Fig.2(b) are included. These are shown by dotted lines in Figs.4(a) and 4(b). We find that the reduction at higher energies $(E > 1 GeV)$ is around 20$\%$ for neutrinos and 40$\%$ for antineutrinos. It is worth noting that the energy dependence of the reduction due to nuclear medium effects is different for neutrinos and antineutrinos. This is due to the different renormalization of various terms like $F_A^2$, $F_2F_A$ and $F_2^2$ in $|T|^2$ which enter in different combinations for neutrino and antineutrino reactions. The results for $\nu_e$ and ${\bar\nu}_e$ cross section are respectively similar to $\nu_\mu$ and ${\bar{\nu}}_\mu$ reactions except for the threshold effects and are not shown here.

In Fig.5 and Fig.6 we compare our results for $\sigma(E)$ with the results of some earlier experiments which contain nuclear targets like Carbon\cite{lsnd}, Freon\cite{bonnetti}-\cite{brunner}, Freon-Propane\cite{pohl} and Aluminum\cite{belikov}, where the experimental results for the deuteron targets\cite{baker} are not included as they are not subject to the various nuclear effects discussed here. It should be kept in mind that the nuclear targets considered here (except for $Br$ in Freon) are lighter than $Fe$. Therefore, the reduction in the total cross section due to nuclear effects will be slightly overestimated. For example, for energies $E_\nu \ge 1GeV$ the reduction in neutrino(antineutrino) cross section in case of $^{56}{Fe}$ is $5\%$ more than the reduction in case of $^{12}{C}$\cite{athar2}. In comparision to the neutrino(antineutrino) nuclear cross sections as obtained in the Fermi gas model of Llewellyn Smith\cite{smith1} (shown by dashed lines in Fig.5 and Fig.6) we get a smaller result for these cross sections. This reduction in the total cross section leads to an improved agreement with the experimental results as compared to the Fermi gas model results specially for antineutrino reactions(Fig.6). It should be emphasized that the Fermi gas model has no specific mechanism to estimate the renormalization of weak transition strengths in nuclei while in our model this is incorporated by taking into account the RPA correlations. 

 In Fig.7(a) and Fig.7(b), we show the nuclear medium effects on the momentum and angular distributions i.e. $\frac{d\sigma}{dp_l}$ and $\frac{d\sigma}{dcos\theta_l}$ of leptons produced in $\nu_\mu$ and ${\bar\nu}_\mu$ reactions. We find a large suppression in the results specially in the peak region of momentum and angular distributions. Quantitatively similar results are obtained for the case of $\nu_e$ and ${\bar\nu}_e$ reactions and are not shown here.

We have also studied the effect of Coulomb distortion in the momentum distribution of leptons but find no substantial effect around $E_\nu=1.0GeV$. These are found to affect the results only at low energies i.e. $E_\nu < 500MeV$ where the peak is slightly shifted to lower momentum as shown in Figs.8(a) and 8(b).
\begin{figure}[h]
\includegraphics{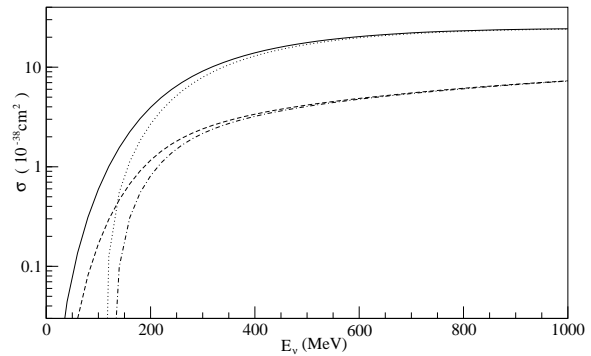}
\caption{Total quasielastic cross sections in the present model for the neutrino and antineutrino reactions in $^{56}{Fe}$ are shown by solid line for $\nu_e$, dashed line for ${\bar{\nu}}_e$, dotted line for $\nu_\mu$  and dashed-dotted line for ${\bar{\nu}}_\mu$ reactions.}
\end{figure}
\begin{figure*}
\includegraphics{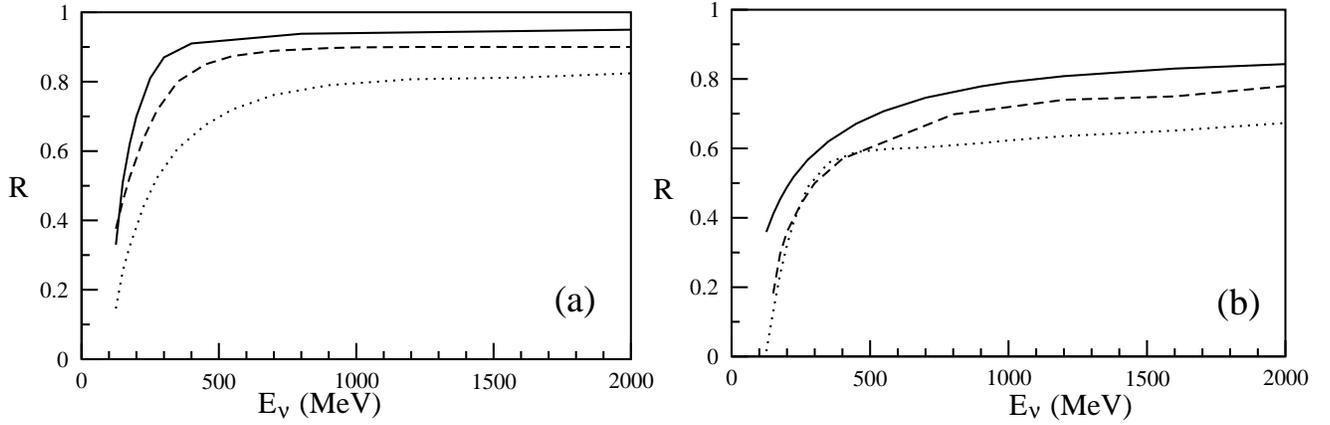}
\caption{Ratio of the total cross section to the free neutrino nucleon cross section for the reactions (a) $\nu_\mu+n\rightarrow \mu^-+p$ (b) ${\bar{\nu}}_\mu+p\rightarrow \mu^++n$ in iron nuclei in the present model with Pauli suppression (solid line) and with nuclear effects (dotted line) and in the Fermi gas model (dashed line)\cite{smith1}.}
\end{figure*}
\begin{figure*}
\includegraphics{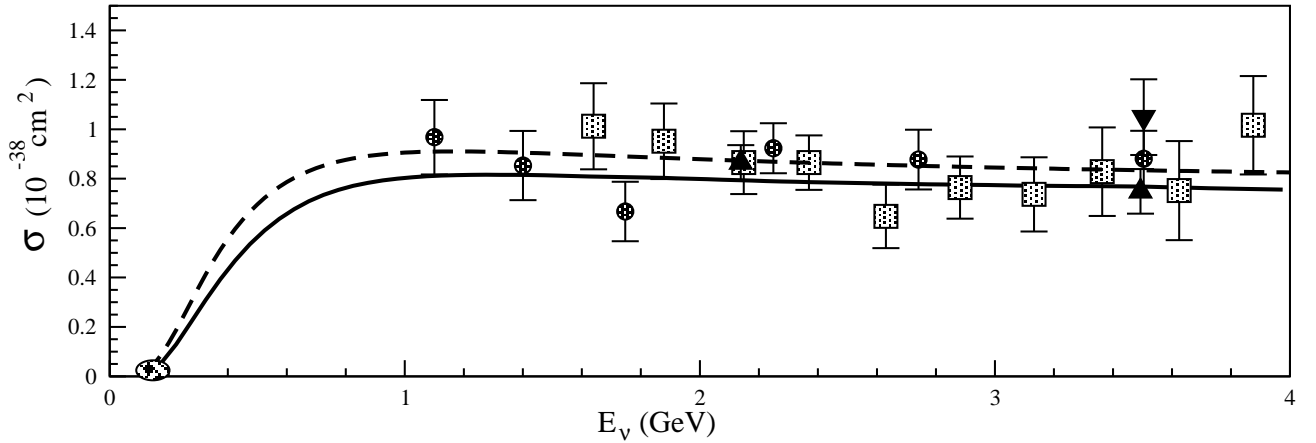}
\caption{Neutrino quasielastic total cross section per nucleon in iron for $\nu_\mu+n\rightarrow p+\mu^-$ reaction. The data are from LSND\cite{lsnd}(Ellipse), Bonnetti et al.\cite{bonnetti}(squares), SKAT collab.\cite{brunner}(Triangle Down), Pohl et al.\cite{pohl}(Circle) and Belikov et al.\cite{belikov}(Triangle Up). The dashed line is the result of the cross section in the Fermi gas model\cite{smith1} and solid line is the result using the present model with nuclear effects.}
\end{figure*}
\begin{figure*}
\includegraphics{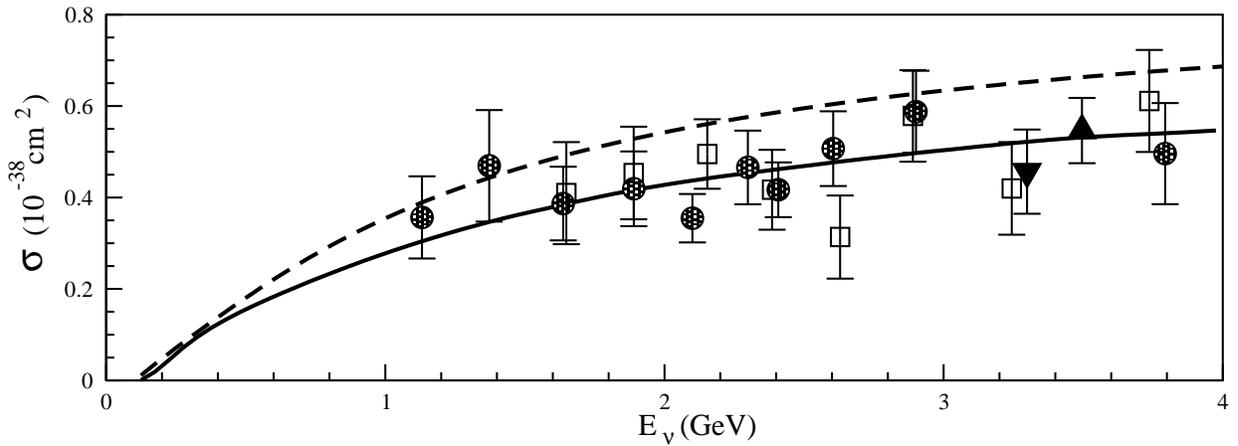}
\caption{Anti-neutrino quasielastic total cross section per nucleon in iron for ${\bar{\nu_\mu}}+p\rightarrow n+\mu^+$ reaction. The data are from Bonnetti et al.\cite{bonnetti}, SKAT collabn.\cite{brunner}(Triangle Down), Pohl et al.\cite{pohl}(Circle), Belikov et al.\cite{belikov}(Triangle Up). The dashed line is the result of the cross section in the Fermi gas model\cite{smith1} and solid line is the result using the present model with nuclear effects.}
\end{figure*}

\begin{figure*}
\includegraphics{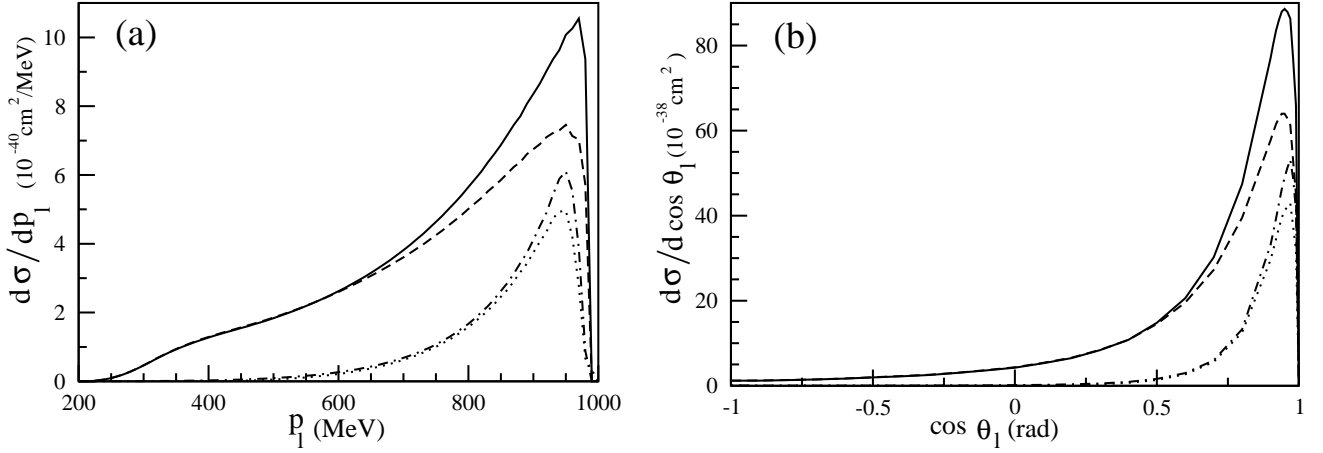}
\caption{The results for the differential cross section Fig.(a) $\frac{d\sigma}{dp_l}$ vs $p_l$ and Fig.(b) $\frac{d\sigma}{d cos\theta_l}$ vs $cos\theta_l$ at $E=1.0GeV$ in the present model with Pauli suppression (solid line for $\nu_\mu$ and dashed-dotted line for $\bar\nu_\mu$) and with nuclear effects (dashed line for $\nu_\mu$ and dotted line for $\bar\nu_\mu$).}
\end{figure*}
\begin{figure*}
\includegraphics{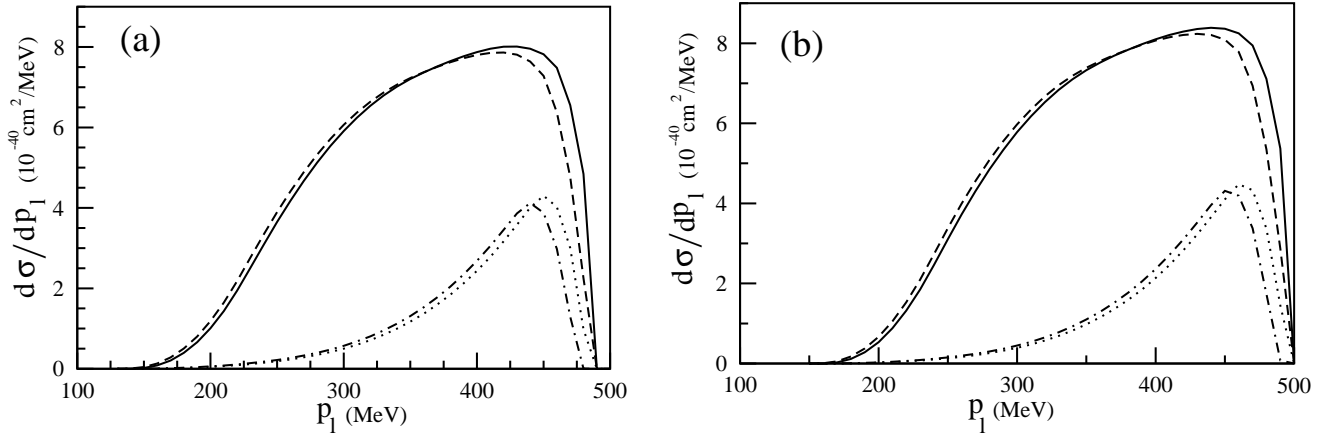}
\caption{Differential scattering cross section $\frac{d\sigma}{dp_l}$ vs $p_l$ for $E_\nu=500MeV$, with Coulomb effect (solid line for $\nu_l$ and dashed-dotted line for $\bar\nu_l$) and without Coulomb effect (dashed line for $\nu_l$ and dotted line for $\bar\nu_l$) Coulomb effect. Fig.(a) is for electron type and Fig.(b) for muon type neutrinos.}
\end{figure*}
\section{Inelastic Production of Leptons}
The inelastic production process of leptons is the process in which the production of leptons is accompanied by one pion (or more pions). There are many calculations of one pion production by neutrinos from free nucleons\cite{adler} but there are only a few calculations which discuss the nuclear effects in these processes\cite{adler1}-\cite{paschos}. In this section we follow the method of Ref.\cite{ruso} to estimate the nuclear effects and nuclear model dependence of inelastic production cross section of leptons induced by neutrinos from iron nuclei. The calculations are done assuming $\Delta$-dominance of one pion production because the contribution of higher resonances in the energy region of atmospheric neutrinos leading to fully contained events is expected to be small.

\subsection{Formalism}
The matrix element for neutrino production reaction of $\Delta$ on proton targets leading to one pion events i.e. 
\begin{equation}
\nu_l(k)+p(p)\rightarrow l^{-}(k^\prime)+\Delta^{++}(p^\prime)
\end{equation}
 is given by Eqn.5 where $J^\mu$ now defines the matrix element of the transition hadronic current between $N$ and $\Delta$ states. The most general form of $J_{CC}^\mu$ is written as\cite{ruso}:
\begin{eqnarray}
J_{cc}^\mu&=&\bar{\psi}_\alpha(p^\prime)\left[\left(\frac{C^V_{3}(q^2)}{M}(g^{\alpha\mu}{\not q}-q^\alpha{\gamma^\mu})\right.\right.\nonumber\\
&+&\frac{C^V_{4}(q^2)}{M^2}(g^{\alpha\mu}q\cdot{p^\prime}-q^\alpha{p^{\prime\mu}})\nonumber\\
&+&\left.\left.\frac{C^V_5(q^2)}{M^2}(g^{\alpha\mu}q\cdot p-q^\alpha{p^\mu})+\frac{C^V_6(q^2)}{M^2}q^\alpha q^\mu\right)\gamma_5\right.\nonumber\\ 
&+&\left.\left(\frac{C^A_{3}(q^2)}{M}(g^{\alpha\mu}{\not q}-q^\alpha{\gamma^\mu})+\frac{C^A_{4}(q^2)}{M^2}(g^{\alpha\mu}q\cdot{p^\prime}-q^\alpha{p^{\prime\mu}})\right.\right.\nonumber\\ 
&+&\left.\left.C^A_{5}(q^2)g^{\alpha\mu}+\frac{C^A_6(q^2)}{M^2}q^\alpha q^\mu\right)\right]u(p)\nonumber\\
\end{eqnarray}
where ${\psi_\alpha}(p^\prime)$ and u(p) are the Rarita Schwinger and Dirac spinors for $\Delta$ and nucleon of momenta $p^\prime$ and $p$ respectively, $q(=p^\prime-p=k-k^\prime)$ is the momentum transfer and $C^V_i$(i=3-6) are vector and $C^A_i$(i=3-6) are axial vector transition form factors. The vector form factors $C^V_i$(i=3-6) are determined by using the conserved vector current(CVC) hypothesis which gives $C_6^V(q^2)$=0 and relates $C_i^V$(i=3,4,5) to the electromagnetic form factors which are determined from photoproduction and electroproduction of $\Delta$'s. Using the analysis of these experiments\cite{paschos}-\cite{dufner} we take for the vector form factors
\begin{eqnarray}
C_{5}^V=0,\;\;\;\;C_{4}^V=-\frac{M}{M_\Delta}C_{3}^V,~~~~~~~~~\mbox{and}\nonumber\\
C_{3}^V(q^2)=\frac{2.05}{(1-\frac{q^2}{M_V^2})^2}, ~M_V^2=0.54GeV^2
\end{eqnarray}
The axial vector form factor $C_{6}^A(q^2)$ is related to ${C_5}^A(q^2)$ using PCAC and is given by
\begin{equation}
C_{6}^A(q^2)={C_5}^A(q^2)\frac{M^2}{{m_\pi}^2-q^2}
\end{equation}
The remaining axial vector form factor ${C^{A}_{i=3,4,5}}(q^2)$ are taken from the experimental analysis of the neutrino experiments producing $\Delta$'s in proton and deuteron targets\cite{kitagaki}-\cite{barish}. These form factors are not uniquely determined but the following parameterizations give a satisfactory fit to the data.
\begin{equation}
{C^{A}_{i=3,4,5}}(q^2)=C_{i}^A(0)\left[1+\frac{{a_i}q^2}{b_i-q^2}\right]\left(1-\frac{q^2}{{M_A}^2}\right)^{-2}
\end{equation}
with $C_{3}^A(0)=0, C_{4}^A(0)=-0.3,C_{5}^A(0)=1.2, a_4=a_5=-1.21,\;\; b_4=b_5=2GeV^2$, $M_A=1.28GeV$. Using the hadronic current given in Eqn.(15), the energy spectrum of the outgoing leptons is given by
\begin{equation}
\frac{d^2\sigma}{dE_{k^\prime}d\Omega_{k^\prime}}=\frac{1}{8\pi^3}\frac{1}{MM^\prime}\frac{k^\prime}{E_\nu}\frac{\frac{\Gamma(W)}{2}}{(W-M^\prime)^2+\frac{\Gamma^2(W)}{4.}}L_{\mu\nu}J^{\mu\nu}  
\end{equation}
where $W=\sqrt{(p+q)^2}$ and $M^\prime$ is mass of $\Delta$,
\[L_{\mu\nu}=k_\mu k_\nu^\prime+k_\mu^\prime k_\nu-g_{\mu\nu}k\cdot k^\prime+i\epsilon_{\mu\nu\alpha\beta}k^\alpha k^{\prime\beta},\]
\[J^{\mu\nu}=\bar{\Sigma}\Sigma J^{\mu\dagger} J^\nu\]
and is calculated with the use of spin $\frac{3}{2}$ projection operator $P^{\mu\nu}$ defined as \[P^{\mu\nu}=\sum_{spins}\psi^\mu {\bar{\psi^\nu}}\] and given by:
\begin{eqnarray}
P^{\mu\nu}=-\frac{\not{p^\prime}+M^\prime}{2M^\prime}\left(g^{\mu\nu}-\frac{2}{3}\frac{p^{\prime\mu} p^{\prime\nu}}{M^{\prime 2}}\right.\nonumber\\
\left.+\frac{1}{3}\frac{p^{\prime\mu} \gamma^\nu-p^{\prime\nu} \gamma_\mu}{M^{\prime}}-\frac{1}{3}\gamma^\mu\gamma^\nu\right)
\end{eqnarray}
In Eqn.(19), the decay width $\Gamma$ is taken to be an energy dependent P-wave decay width given by
\begin{eqnarray}
\Gamma(W)=\frac{1}{6\pi}\left(\frac{f_{\pi N\Delta}}{m_\pi}\right)^2\frac{M}{W}|{{\bf q}_{cm}|^3}\Theta(W-M-m_\pi)
\end{eqnarray}
where 
\[|{\bf q}_{cm}|=\frac{\sqrt{(W^2-m_\pi^2-M^2)^2-4m_\pi^2M^2}}{2W}\]
and $M$ is the mass of nucleon. The step function $\Theta$ denotes the fact that the width is zero for the invariant masses below the $N\pi$ threshold. ${|\bf q_{cm}|}$ is the pion momentum in the rest frame of the resonance. When the reaction(14) takes place in the nucleus, the neutrino interacts with the nucleon moving inside the nucleus of density $\rho(r)$ with its corresponding momentum $\vec{p}$ constrained to be below its Fermi momentum. The produced $\Delta$'s have no such constraints on their momentum. These $\Delta's$ decay through various decay channels in the medium. The most prominent decay mode is $\Delta\rightarrow N\pi$ which produces pions. This decay mode in the nuclear medium is slightly inhibited due to Pauli blocking of the final nucleon momentum modifying the decay width $\Gamma$ used in Eqn.(21). This modification of $\Gamma$ due to Pauli blocking of nucleus has been studied in detail in electromagnetic and strong interactions\cite{oset}. The modified $\Delta$ decay width i.e. $\tilde\Gamma$ is written as\cite{oset}: 
\begin{equation}
\tilde\Gamma=\frac{1}{6\pi}\left (\frac{f_{\pi N \Delta}}{m_{\pi}}\right )^{2}\frac{M|{\bf q}_{cm}|^{3}}{W} F(k_{F},E_{\Delta},k_{\Delta})\Theta(W-M-m_\pi) 
\end{equation}
where $F(k_{F},E_{\Delta},k_{\Delta})$ is the Pauli correction factor given by 
\begin{equation}
F(k_{F},E_{\Delta},k_{\Delta})= \frac{k_{\Delta}|{{\bf q}_{cm}}|+E_{\Delta}{E^\prime_p}_{cm}-E_{F}{W}}{2k_{\Delta}|{\bf q^\prime}_{cm}|} 
\end{equation}
where $k_{F}$ is the Fermi momentum, $E_F=\sqrt{M^2+k_F^2}$, $k_{\Delta}$ is the $\Delta$ momentum and  $E_\Delta=\sqrt{W+k_\Delta^2}$.

Moreover, in the nuclear medium there are additional decay channels now open due to two body and three body absorption processes like $\Delta N \rightarrow N N$ and $\Delta N N\rightarrow N N N$ through which $\Delta's$ disappear in the nuclear medium without producing a pion while a two body $\Delta$ absorption process like $\Delta N  \rightarrow \pi N N$ gives rise to some more pions. These nuclear medium effects on $\Delta$ propagation are included by using a $\Delta$ propagator in which the $\Delta$ propagator is written in terms of $\Delta$ self energy $\Sigma_\Delta$. This is done by using a modified mass and width of $\Delta$ in nuclear medium i.e.  $M_\Delta \rightarrow M_\Delta + Re\Sigma_\Delta$ and $\tilde\Gamma \rightarrow \tilde\Gamma - Im\Sigma_\Delta$. There are many calculations of $\Delta$ self energy $\Sigma_\Delta$ in the nuclear medium \cite{oset}-\cite{oset3} and we use the results of \cite{oset}, where the density dependence of real and imaginary parts of $\Sigma_\Delta$ are parametrized in the following form:
\[Re{\Sigma}_{\Delta}=40 \frac{\rho}{\rho_{0}}MeV ~~\mbox{and} \, \hspace{70mm} \]
\begin{equation}
-Im{{\Sigma}_{\Delta}}=C_{Q}\left (\frac{\rho}{{\rho}_{0}}\right )^{\alpha}+C_{A2}\left (\frac{\rho}{{\rho}_{0}}\right )^{\beta}+C_{A3}\left (\frac{\rho}{{\rho}_{0}}\right )^{\gamma}
\end{equation}
In Eqn.24, the term with $C_{Q}$ accounts for the $\Delta N  \rightarrow \pi N N$ process, the term with $C_{A2}$ for two-body absorption process $\Delta N \rightarrow N N$ and the term with $C_{A3}$ for three-body absorption process $\Delta N N\rightarrow N N N$. The coefficients $C_{Q}$, $C_{A2}$, $C_{A3}$, $\alpha$, $\beta$ and $\gamma$ ($\gamma=2\beta$) are parametrized in the range $80<T_{\pi}<320MeV$ (where $T_{\pi}$ is the pion kinetic energy) as \cite{oset}
\begin{eqnarray}
C_i(T_{\pi})&=&ax^{2}+bx+c,~ \mbox{for}~ x=\frac{T_{\pi}}{m_{\pi}} 
\end{eqnarray}
The values of coefficients $a$, $b$ and $c$ are given in Table-I. taken from ref.\cite{oset}.
\begin{table}[h]
\caption{Coefficients of Eqn.(25) for an analytical interpolation of $Im\Sigma_\Delta$}
\begin{ruledtabular}
\begin{tabular}{cccccc}
&$C_Q$(MeV) & $C_{A2}$(MeV)& $C_{A3}$(MeV) & $\alpha$&$\beta$ \\ \hline
 a & -5.19 & 1.06&-13.46&0.382&-0.038 \\
 b & 15.35&-6.64&46.17&-1.322&0.204\\
 c & 2.06&22.66&-20.34&1.466&0.613 \\
\end{tabular}
\end{ruledtabular}
\end{table}

With these modifications, which incorporate the various nuclear medium effects on $\Delta$ propagation, the cross section is now written as

\begin{eqnarray}
\sigma&=&\int \int \frac{d{\bf r}}{8\pi^3}\frac{d\bf{k^\prime}}{E_\nu E_l}\frac{1}{MM^\prime} \nonumber\\
&&\times \frac{\frac{\tilde\Gamma}{2}-Im\Sigma_\Delta}{(W- M^\prime-Re\Sigma_\Delta)^2+(\frac{\tilde\Gamma}{2.}-Im\Sigma_\Delta)^2}\nonumber\\
&&\times\left[\rho_p({\bf r})+\frac{1}{3}\rho_n({\bf r})\right]L_{\mu\nu}J^{\mu\nu}
\end{eqnarray}
The factor $\frac{1}{3}$ in front of $\rho_n$ comes due to suppression of charged pion production from neutron targets i.e. $\nu_l + n\rightarrow l^- + \Delta^{+} \rightarrow l^- + n + \pi^+$, as compared to the charged pion production from the proton target i.e.$\nu_l + p\rightarrow l^- + \Delta^{++} \rightarrow l^- + p + \pi^+$, in the nucleus. In case of antineutrino reactions $\rho_p + \frac{1}{3}\rho_n$ is replaced by $\rho_n + \frac{1}{3}\rho_p$.

\subsection{Results}
In this section we present results of inelastic lepton production cross section due to $\Delta$h excitations in iron induced by the charged current neutrino interactions using Eqn.(26). In Fig.9(a) and Fig.9(b), we show the results $\sigma(E_\nu)\sim E_\nu$ for $\nu_e({\bar{\nu}}_e)$ and $\nu_\mu({\bar\nu}_\mu)$ for the inelastic production of lepton accompanied with $\Delta$ and compare this with the cross section for the quasielastic production of leptons discussed in section-II. We see that for $E_\nu \approx 1.4GeV$ the lepton production cross section through quasielastic and inelastic production processes are comparable. For energies $E_\nu\le 1.4GeV$ where the atmospheric neutrino energies are important for fully contained events the major contribution comes from the quasielastic events. 

The effect of nuclear effects on the $\Delta$ production are shown in Fig.10 for $\nu_\mu$ and ${\bar\nu}_\mu$ reactions. The results for the $\nu_e$ and ${\bar\nu}_e$ are similar to Fig.10 and are not shown here. We see that the nuclear medium effects reduce the $\Delta$ production cross section by 5-10$\%$.

In Fig.11(a) and Fig.11(b), we show the momentum distribution $d\sigma/dp$ and angular distribution $d\sigma/dcos\theta$ for muon type neutrinos where we also show the effects of nuclear effects. The nuclear medium effects reduce the cross sections mainly in the peak region of the momentum and angular distributions for muons. The results for the momentum distribution and angular distribution for electrons is similar to the muon distributions. We show our final results on momentum and angular distributions for all charged leptons i.e. $e^-, e^+, \mu^-$ and $\mu^+$ produced in neutrino and antineutrino reactions with $\nu_e$, $\nu_\mu$, $\bar\nu_e$ and $\bar\nu_\mu$ for $E_\nu=1GeV$ in Fig.12(a) and Fig.12(b) in $^{56}Fe$ nuclei.

Here we will like to make some comments about the lepton events produced through the $\Delta$ excitation. The $\Delta$ excitation process gives rise to leptons accompanied by one pion events produced by $\Delta \rightarrow N\pi$ and $\Delta N \rightarrow \pi NN$ processes in the nuclear medium. The pions produced through these processes will undergo secondary nuclear interactions like multiple scattering and possible absorption in iron nuclei while passing through the nucleus and an appropriate model has to be used for their description. Models developed by Salcedo et al.\cite{salcedo} and Paschos et al.\cite{paschos} have studied these effects but we do not consider these effects here as they are beyond the scope of this paper. The $\Delta$ excitation process in the nuclear medium also gives rise to quasielastic like events through two body and three body absorption processes like $\Delta N\rightarrow NN$ and $\Delta N\rightarrow \Delta NN$ where only leptons are present in the final states. The quasielastic-like lepton events are discussed by Kim et al.\cite{kim} but no quantitative estimates have been made. We have in an earlier paper\cite{ruso} discussed this process only qualitatively but make a quantitative estimate of these events in this paper. We find that around $E_\nu=1GeV$ the contribution of these quasielastic like events is about $10-12\%$ but its effect on the flux averaged production of leptons for atmospheric neutrinos is not very significant. This is discussed in some detail in the next section. 
\section{Lepton Production by Atmospheric neutrinos}
The energy dependences of the quasielastic and inelastic lepton production cross sections described in sections II and III have been used to calculate the lepton production by atmospheric neutrinos after averaging over the neutrino flux corresponding to the two sites of Soudan and Gransasso, where iron based detectors are being used.
There are quite a few calculations of atmospheric neutrino fluxes at these two sites. We use the angle averaged fluxes calculated by Honda et al.\cite{honda} and Barr et al.\cite{barr} for the Soudan site and the fluxes of Barr et al.\cite{barr} and Plyaskin\cite{plyaskin} for the Gransasso site to calculate the flux averaged cross section $<\sigma>$ and also the momentum and the angular distributions $<\frac{d\sigma}{dp_l}>$ and $<\frac{d\sigma}{d cos\theta_l}>$ for leptons produced by $\nu_e$, $\bar\nu_e$, $\nu_\mu$ and $\bar\nu_\mu$.

\subsection{Flux averaged Momentum and Angular distributions}
In this section we present the numerical results for the flux averaged momentum distributions $<\frac{d\sigma}{dp_l}>$ and angular distributions $<\frac{d\sigma}{dcos\theta_l}>$ for various leptons produced from $\nu_e$, ${\bar{\nu}}_e$, $\nu_\mu$, ${\bar{\nu}}_\mu$ at the atmospheric neutrino sites of Soudan and Gransasso. These leptons are produced through the quasielastic as well as inelastic processes. The quasielastic production has been discussed in section-II and the inelastic production has been discussed in section-III. The inelastic production of leptons is accompanied by pions. However, in the nuclear medium where the production of $\Delta$ is followed by $\Delta N \rightarrow NN$, $\Delta NN \rightarrow NNN$ absorption processes, the leptons are produced without pions. These are quasielastic like events. We show the momentum distribution of the leptons produced in iron nuclei corresponding to one pion and quasielastic like events for atmospheric neutrinos relevant to Soudan and Gransasso sites. We show it for $\nu_\mu$ and $\bar\nu_\mu$ for the Soudan site in Figs. 13(a) and 13(b) and for the Gransasso site in Figs.14(a) and 14(b) corresponding to the flux of Barr et al.\cite{barr}. We see that at both sites, the production cross section of quasielastic-like events is quite small.

We now present our final results for the momentum distribution and angular distribution for various leptons produced by atmospheric neutrinos at Soudan and Gransasso sites in Figs. 15-18. In these figures separate contributions from the quasielastic and inelastic processes to the momentum and angular distributions of charged leptons are shown explicitly. The quasielastic events are those where only a charge lepton is produced in the final state either by a quasielastic process described in section-II or by an inelastic process of $\Delta$-production followed by its subsequent absorption in the nuclear medium as described in section-III. The inelastic events are those events in which a charged lepton in the final state is accompanied by a charged pion as a decay product of deltas excited in the nuclear medium.

We show the results for the momentum distribution $<d\sigma/dp_l>\sim p_l(l=e^-,e^+,\mu^+,\mu^-)$ for the atmospheric neutrino fluxes of Barr et al.\cite{barr} at Soudan site in Fig.15 and at Gransasso site in Fig.16. We see that in all cases the major contributions to the charged lepton production comes from the quasielastic processes induced by neutrinos(solid line). The contribution from the quasielastic processes induced by antineutrinos(dotted line) and inelastic processes induced by neutrinos(dashed line) is small and is around $20\%$ in the peak region. The contribution due to inelastic processes induced by antineutrinos is very small over the whole region (dashed-dotted lines).

The peak in quasielastic $\nu_\mu$ reactions occurs around $p_l\approx 200MeV$. The peaks in the inelastic $\nu_\mu$ and quasielastic ${\bar{\nu_\mu}}$ reactions are slightly shifted towards lower energies. The momentum distribution of the leptons for the quasielastic reaction is peaked more sharply than the momentum distribution of the inelastic reaction. 

We have also presented the numerical results for angular distributions of leptons $<\frac{d\sigma}{dcos\theta_l}> ~vs~ cos\theta$ for the atmospheric neutrino fluxes of Barr et al.\cite{barr} for the Soudan site in Fig.17 and for the Gransasso site in Fig.18. The inelastic lepton production accompanied by pions (dashed line for $\mu^-(e^-)$ production and dashed-dotted line for $\mu^+(e^+)$ production) and the quasielastic lepton production events (solid line for $\mu^-(e^-)$ production and dotted line for $\mu^+(e^+)$ production) have been explicitly shown in these figures. We see from these figures that the lepton production cross sections are forward peaked in all cases. The inelastic distribution due to pion production are slightly more forward peaked than the quasielastic distribution. The contribution of the inelastic lepton events are small compared to the quasielastic events and the angular distribution of the flux averaged cross sections are dominated by the quasielastic events.

At Soudan site, we have also studied the momentum and angular distributions for the atmospheric neutrino fluxes of Honda et al.\cite{honda}. We find that the momentum and angular distributions for the production of muons are similar to the distributions obtained for the flux of Barr et al.\cite{barr}. In the case of electron production, the use of the flux of Honda et al.\cite{honda}, gives a slightly smaller value for $<d\sigma/dp_l>$ and $<\frac{d\sigma}{dcos\theta_l}>$ for electrons as compared to the flux of Barr et al.\cite{barr}.

Similarly at Gransasso site, the momentum and angular distributions for the atmospheric neutrino fluxes of Plyaskin\cite{plyaskin} have also been studied. Here, also we find that the momentum and angular distributions for the production of muons are similar to the distributions obtained for the flux of Barr et al.\cite{barr}. In the case of electron production, the use of the flux of Plyaskin\cite{plyaskin}, gives a slightly smaller value for $<d\sigma/dp_l>$ and $<\frac{d\sigma}{dcos\theta_l}>$ for electrons as compared to the flux of Barr et al.\cite{barr}. 

We further find that for the flux of Barr et al.\cite{barr}, the lepton production cross section at the Soudan site is slightly larger than the Gransasso site for muons as well as for electrons. 
\subsection{Flux averaged Total cross sections and lepton yields}
The total cross sections for the production of leptons and its energy dependence have been discussed in section II and section III for quasielastic and inelastic reactions. In this section we calculate the lepton yields $Y_l$ for lepton of flavor $l$ which we define as
\[Y_l=\int \Phi_{\nu_l}~\sigma(E_{\nu_l})~dE_{\nu_l}\]
where, $\Phi_{\nu_l}$ is the atmospheric neutrino flux of $\nu_l$ and $\sigma(E_{\nu_l})$ is the total cross section for neutrino $\nu_l$ of energy $E_{\nu_l}$. We calculate this yield separately for the quasielastic and inelastic events. We define a relative yield of muon over electron type events by $R$ as $R=R_{\mu/e}=\frac{Y_\mu +Y_{\bar\mu}}{Y_e+Y_{\bar e}}$ for quasielastic and inelastic events and present our results in Table-II. We study the nuclear model dependence as well as the flux dependence of the relative yield $R$. 

The results for $R$ are presented separately for quasielastic events, inelastic events and the total events in Table-II. For quasielastic events $\nu_l({\bar{\nu_l}})+^{56}{Fe}\rightarrow l^-(l^+)+X$, the results are presented for the case of free nucleon by $R_{FN}$, for the nuclear case with Fermi gas model description of Llewellyn Smith\cite{smith1} by $R_{FG}$ and for the case of nuclear effects within our model by $R_{NM}$. We see that there is practically no nuclear model dependence on the value of $R$ (compare the values of $R_{NM}$, $R_{FG}$ and $R_{FN}$ for the same fluxes at each site). This is also true for the inelastic production of leptons i.e. $\nu_l({\bar{\nu_l}})+^{56}{Fe}\rightarrow l^-(l^+)+\pi^+(\pi^-)+X$ for which the ratio for free nucleon case (denoted by $R_{\Delta F}$) and the ratio for the nuclear case in our model (denoted by $R_{\Delta N}$) are presented in Table II. It is, therefore, concluded that there is no appreciable nuclear model dependence on the ratio of total lepton yields for the production of muons and electrons(compare the values of $R_F$ and $R_N$, where $R_F$ shows the ratio of total lepton yields for muon and electron for the case of free nucleon and $R_N$ shows the ratio of total yield for muon and electron for the case of nucleon in the nuclear medium). 

However, there is some dependence of the ratio R on the atmospheric neutrino fluxes. The flux dependence of $R$ can be readily seen from Table-II, for the two sites of Soudan and Gransasso. At the Gransasso site, we see that there is 4-5$\%$ difference in the value of $R_N$ for the total lepton yields for the fluxes of Barr et al.\cite{barr} and Plyaskin\cite{plyaskin}. At Soudan site, the results for the fluxes of Honda et al.\cite{honda} and Barr et al.\cite{barr} are within 4-5$\%$ but the flux calculation of Plyaskin\cite{plyaskin} gives a result which is about 10-11$\%$ smaller than the results of Honda et al.\cite{honda} and 7-8$\%$ smaller than the results of Barr et al.\cite{barr}. The flux dependence is mainly due to the quasielastic events. This should be kept in mind while using the flux of Plyaskin\cite{plyaskin} for making any analysis of the neutrino oscillation experiments.

In Table-III, we present a quantitative estimate of the relative yield of inelastic events  $r_l$ defined by $r_l=\frac{{Y_l}^\Delta}{Y_l}$, where ${Y_l}^\Delta={Y_l}^\Delta + {Y_{\bar l}}^\Delta$ is the lepton yield due to the inelastic events and $Y_l$ is the total lepton yield due to the quasielastic and inelastic events i.e.$Y_l={Y_l}^{q.e.} + {Y_{\bar l}}^{q.e.} + {Y_l}^\Delta + {Y_{\bar l}}^\Delta$. The relative yield for the case of free nucleon is shown by $r_l(F)$ and for the case with the nuclear effects in our model is shown by $r_l(N)$. We see that for free nucleons, the relative yield of the inelastic events due to $\Delta$ excitation is in the range of 12-15$\%$ for various fluxes at the two sites. The ratio is approximately same for electrons and muons. When the nuclear effects are taken into account this becomes 19-22$\%$. This is due to different nature of the effect of nuclear structure on the quasielastic and inelastic production cross scetions which gives a larger reduction in the cross section for the quasielastic case as compared to the inelastic case. This quantitative estimate of $r_l(N)$ in iron may be compared with the results in oxygen where the experimental results at Kamiokande give a value of 18$\%$\cite{kajita11}.

\section{Conclusions}
We have studied the charged lepton production in iron induced by atmospheric neutrinos at the experimental sites of Soudan and Gransasso. The energy dependence of the total cross sections for the quasielastic and inelastic processes have been calculated in a nuclear model which takes into account the effect of Pauli principle, Fermi  motion effects and the renormalization of weak transition strengths in nuclear medium. The inelastic process has been studied through the $\Delta$ dominance model which incorporates the modification of mass and width of $\Delta$ resonance in the nuclear medium. The numerical results for the momentum and angular distributions of the charged lepton production cross section have been presented for muons and electrons. The relative yield of muon to electron production has been studied. In addition to the nuclear model dependence, the flux dependence of the total yields, momentum distribution $\frac{d\sigma}{dp_l}$ and $\frac{d\sigma}{dcos\theta_l}$ have been also studied. In the following we conclude this paper by summarizing our main results:

1.There is a large reduction due to nuclear effects in the total cross section for quasielastic production cross section specially at lower energies(40-50$\%$ around 200MeV) and the reduction becomes smaller at higher energies(15-20$\%$ around 500MeV ).

2.For quasielastic reactions we find a larger reduction in the total cross section as compared to the Fermi gas model. The energy dependence of this reduction in cross section at low energies ($E < 500MeV$) is found to be different for neutrino and antineutrino reactions.

3.The inelastic production of leptons where a charged lepton is accompanied by a pion becomes comparable to the quasielastic production of leptons for $E_\nu\sim 1.4GeV$ and increases with increase of energy. The effect of nuclear medium on the inelastic production cross section (in absence of pion re-scattering effects) in not too large ($\sim 10\%$).

4. For quasielastic reactions the effect of Coulomb distortion of the final lepton in the total cross section is small except at very low energies ($E < 500MeV$) and becomes negligible when averaged over the flux of atmospheric neutrinos. 

5.The flux averaged momentum distribution of leptons produced by atmospheric neutrinos is peaked around the momentum $p_l\sim 200MeV$ for electrons and muons. The peak for the quasielastic production is sharper than the inelastic production. The inelastic production of leptons contributes about 20$\%$ to the total production of leptons.

6.The flux averaged angular distribution of leptons for atmospheric neutrinos for quasielastic as well as inelastic production is sharply peaked in the forward direction.

7.There is a very little flux dependence on the relative yield of muons and electrons at the site of Gransasso. However, at the Soudan site, the atmospheric flux as determined by Plyaskin\cite{plyaskin} gives a value of relative yield which is smaller than the relative yield using the flux of Honda et al.\cite{honda} and Barr et al.\cite{barr}.

8.The nuclear model dependence of the relative yield of muons to electrons is negligible, even though the individual yields for muons and electrons are reduced with the inclusion of nuclear effects specially for the quasielastic production.
\section{Acknowledgment}
The work is financially supported by the Department of Science and Technology, Government of India under the grant DST Project No. SP/S2K-07/2000. One of the authors (S.Ahmad) would loke to thank CSIR for financial support.
\begin{table*}
\caption{Ratio $R=R_{\mu/e}=\frac{Y_\mu +Y_{\bar\mu}}{Y_e+Y_{\bar e}}$ corresponding to quasielastic, inelastic and total production of leptons [FG refers to Fermi Gas Model, NM refers to Nuclear Model, FN refers to Free Nucleon, $\Delta$N refers to $\Delta$ in Nuclear Model, $\Delta$F refers to $\Delta$ Free]. $R_F$ shows the ratio of total lepton yields for muon to electron for the case of free nucleon and $R_N$ shows the ratio of total yields for muon to electron for the case of nucleon in the nuclear medium.}
\begin{tabular}{|c|c|c|c|c|c|}\hline\hline
  Site &Soudan &Soudan&Soudan &Gransasso&Gransasso\\ \hline
 FLUX &Barr et al.\cite{barr} &Plyaskin\cite{plyaskin}&Honda et al.\cite{honda} &Barr et al.\cite{barr}
&Plyaskin\cite{plyaskin}\\ \hline
Quasielastic & & & & &\\ \hline\hline
$R_{NM}$& 1.80& 1.65 &1.89 &1.95 & 2.00 \\
$R_{FG}$& 1.81& 1.66 &1.89 &1.95 & 2.09 \\
$R_{FN}$& 1.82& 1.68 &1.90 &1.95 & 2.08 \\ \hline
Inelastic & & & & &\\ \hline\hline
$R_{\Delta N}$& 1.84& 1.81 &1.95 &2.02 & 2.05 \\
$R_{\Delta F}$& 1.82& 1.80 &1.94 &2.01 & 2.03 \\ \hline
Total & & & & &\\ \hline\hline
$R_{N}$& 1.81& 1.68 &1.90 &1.96 & 2.01 \\ 
$R_{F}$& 1.82& 1.69 &1.90 &1.96 & 2.07 \\ \hline\hline
\end{tabular}
\end{table*}

\begin{table*}
\caption{\% ratio  $r=\frac{Y_\Delta}{Y_{q.e.+ \Delta}}$ [N refers to Nuclear Model, F refers to free case]}
\begin{tabular}{|c|c|c|c|c|c|}\hline\hline
 Sites &Soudan&Soudan &Soudan &Gransasso&Gransasso\\ \hline
 FLUX &Barr et al.\cite{barr}&Plyaskin\cite{plyaskin} &Honda et al.\cite{honda} &Barr et al.\cite{barr}
&Plyaskin\cite{plyaskin}\\ \hline
$r_{\mu}{(F)}$& 14& 12 &14 &15 & 12 \\
$r_{e}{(F)}$& 14& 11 &14 &14 & 12 \\
$r_{\mu}{(N)}$& 22& 20 &22 &23 & 19 \\
$r_{e}{(N)}$& 22& 19 &22 &22 & 19 \\ \hline\hline
\end{tabular}
\end{table*}

\begin{figure*}
\includegraphics{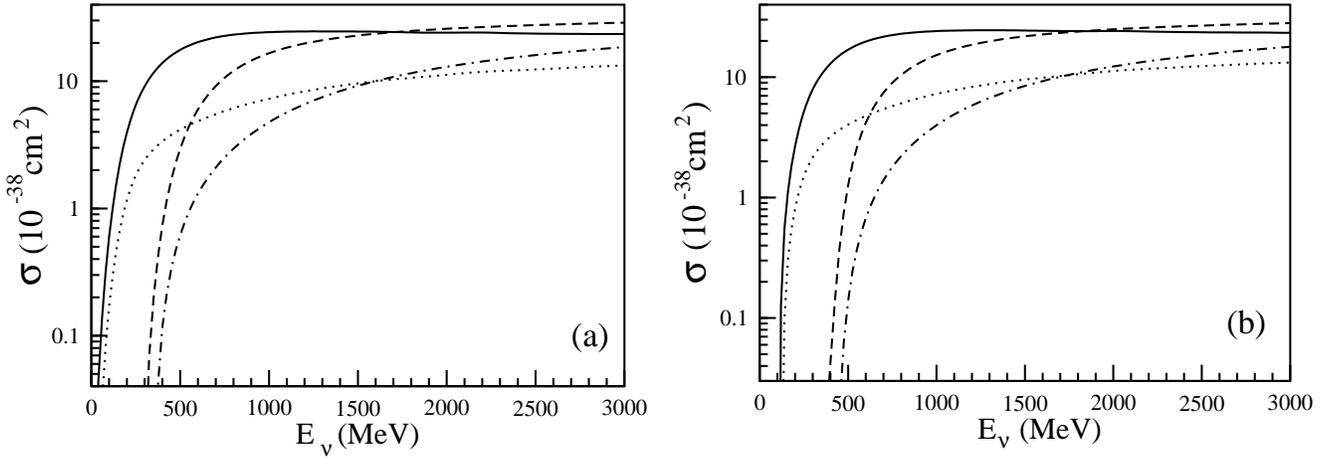}
\caption{The total scattering cross section for the neutrino $\nu_l+N \rightarrow \Delta+l^-$ (shown by dashed line) and anti-neutrino $\bar\nu_l+N \rightarrow \Delta+l^+$ reactions (shown by dashed-dotted line) in $^{56}Fe$ and compared with the corresponding quasielastic cross section in the present model with nuclear effects for the neutrino (solid line) and anti-neutrino(dotted line) reactions in $^{56}Fe$. Fig.(a) is for electron type and Fig.(b) is for muon type neutrinos. }
\end{figure*}

\begin{figure}
\includegraphics{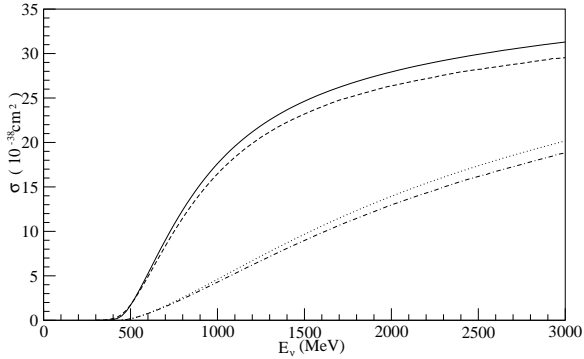}
\caption{Total cross section at $E_\nu =1.0GeV$ for the reaction ${\nu_l}({\bar\nu}_l)+N \rightarrow \Delta+l^-(l^+)$ in $^{56}Fe$ with (dashed line for $\nu_\mu$, dashed-dotted for $\bar\nu_\mu$) and without nuclear effects(solid line is the result for $\nu_\mu$ reaction and dotted for $\bar\nu_\mu$.)}
\end{figure}

\begin{figure*}
\includegraphics{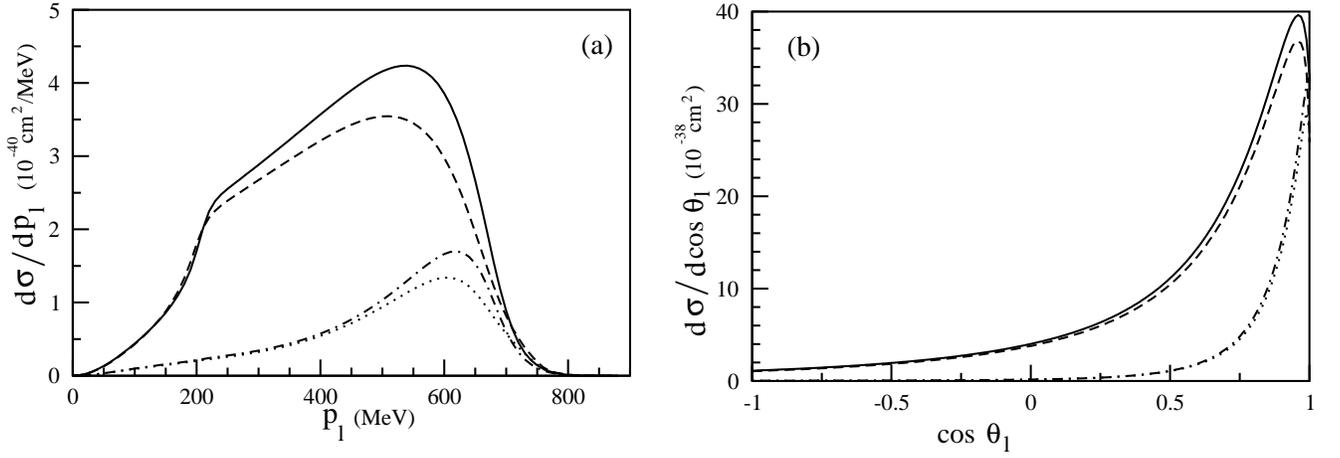}
\caption{Differential scattering cross section Fig.(a) $\frac{d\sigma}{dp_l}$ vs $p_l$ and Fig.(b) $\frac{d\sigma}{dcos\theta_l}$ vs $cos\theta_l$ at $E_\nu =1.0GeV$ for the reaction ${\nu_l}({\bar\nu}_l)+N \rightarrow \Delta+l^-(l^+)$ in $^{56}Fe$ with (dashed line for $\nu_\mu$, dotted for $\bar\nu_\mu$) and without nuclear effects(solid line is the result for $\nu_\mu$ reaction and dashed-dotted for $\bar\nu_\mu$).}
\end{figure*}

\begin{figure*}
\includegraphics{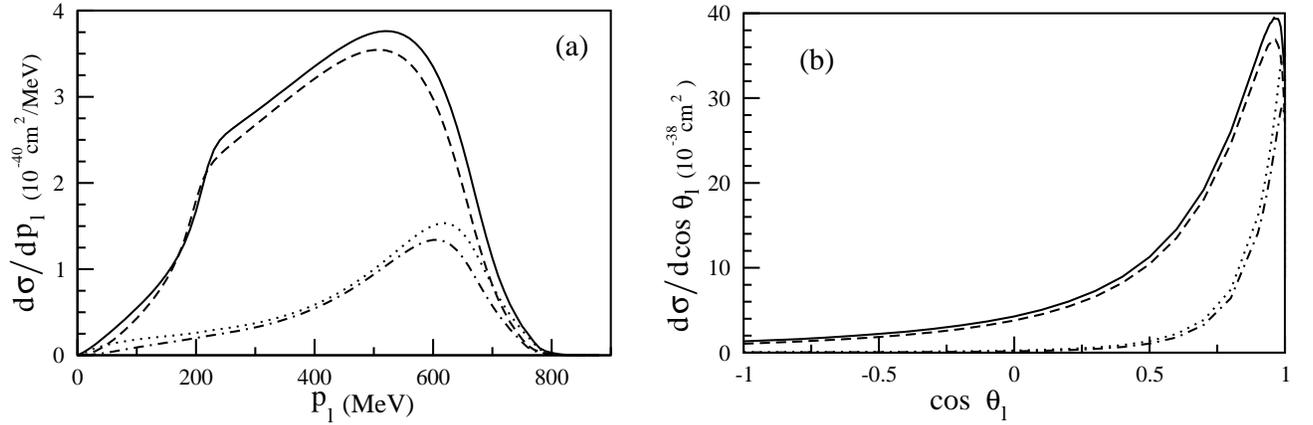}
\caption{The results for the differential cross section Fig.(a) $\frac{d\sigma}{dp_l}$ vs $p_l$ and Fig.(b) $\frac{d\sigma}{dcos\theta_l}$ vs $cos\theta_l$ with nuclear effects at $E=1.0GeV$ for the reaction ${\nu_l}({\bar\nu}_l)+N \rightarrow \Delta+l^-(l^+)$ in $^{56}Fe$. Solid line is the result for $\nu_e$, dashed for $\nu_\mu$, dotted for $\bar\nu_e$ and dashed-dotted for $\bar\nu_\mu$ reactions.}
\end{figure*}

\begin{figure*}
\includegraphics{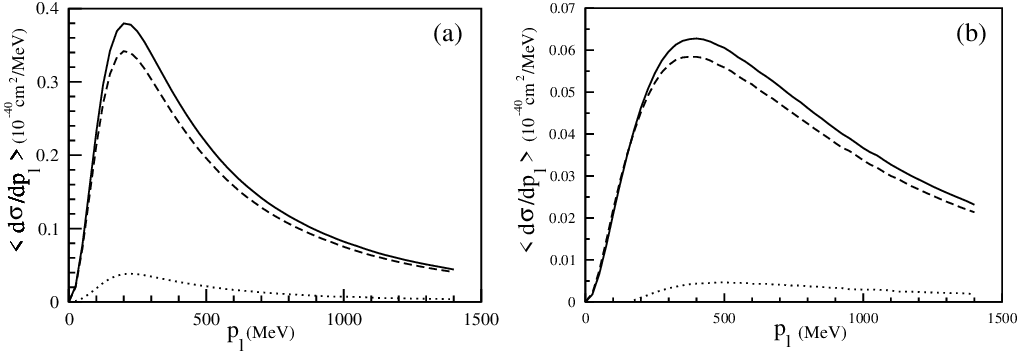}
\caption{Flux averaged differential cross section $<\frac{d\sigma}{dp_l}>$ vs $p_l$ at the Soudan site simulated by Barr et al.\cite{barr}. Fig.(a) for $\nu_\mu + N \rightarrow \Delta + \mu^-$ and Fig.(b) for $\bar\nu_\mu + N \rightarrow \Delta + \mu^+$ reactions in $^{56}Fe$. The solid line is for total $\Delta$ events, dashed line is for one $\pi$ production events and dotted line is for quasielastic like events(due to $\Delta$ absorption in the nuclear medium)}
\end{figure*}

\begin{figure*}
\includegraphics{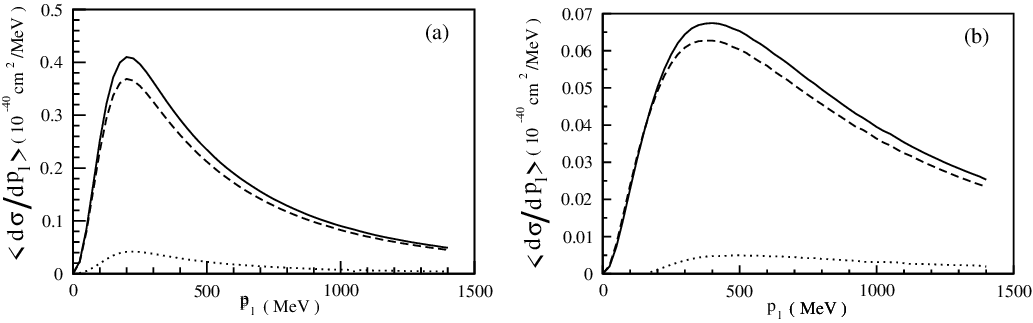}
\caption{Flux averaged differential cross section $<\frac{d\sigma}{dp_l}>$ vs $p_l$ at the Gransasso site simulated by Barr et al.\cite{barr}. Fig.(a) for $\nu_\mu + N \rightarrow \Delta + \mu^-$ and Fig.(b) for $\bar\nu_\mu + N \rightarrow \Delta + \mu^+$ reaction in $^{56}Fe$. The solid line is for total $\Delta$ events, dashed line is for one $\pi$ production events and dotted line is for quasielastic like events(due to $\Delta$ absorption in the nuclear medium)}.
\end{figure*}

\begin{figure*}
\includegraphics{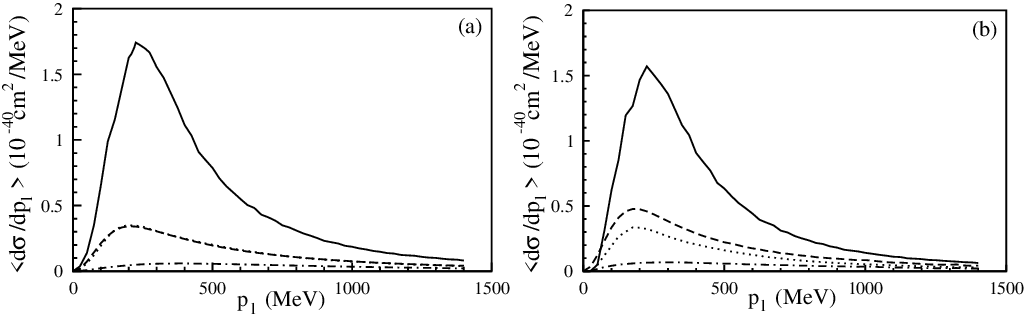}
\caption{Flux averaged differential cross section $<\frac{d\sigma}{dp_l}>$ vs $p_l$ at the Soudan site simulated by Barr et al.\cite{barr}. The solid line is for total quasielastic events and dashed line is the one $\pi$ production events for $\nu_\mu$ reaction in $^{56}{Fe}$. The corresponding results for $\bar\nu_\mu$ are shown by dotted and dashed-dotted lines, Fig.(a) for muon type and Fig.(b) for electron type neutrinos.}
\end{figure*}

\begin{figure*}
\includegraphics{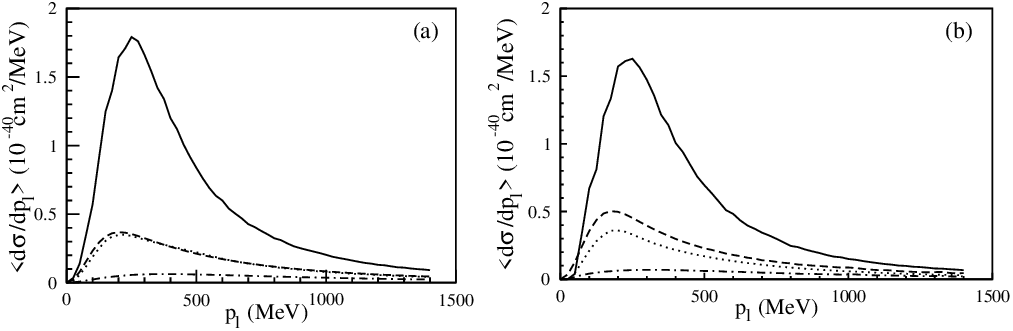}
\caption{Flux averaged differential cross section $<\frac{d\sigma}{dp_l}>$ vs $p_l$ at the Gransasso site simulated by Barr et al.\cite{barr}. The solid line is for total quasielastic events and dashed line is the one $\pi$ production events for $\nu_\mu$ reaction in $^{56}{Fe}$. The corresponding results for $\bar\nu_\mu$ are shown by dotted and dashed-dotted lines, Fig.(a) for muon type and Fig.(b) for electron type neutrinos.}
\end{figure*}

\begin{figure*}
\includegraphics{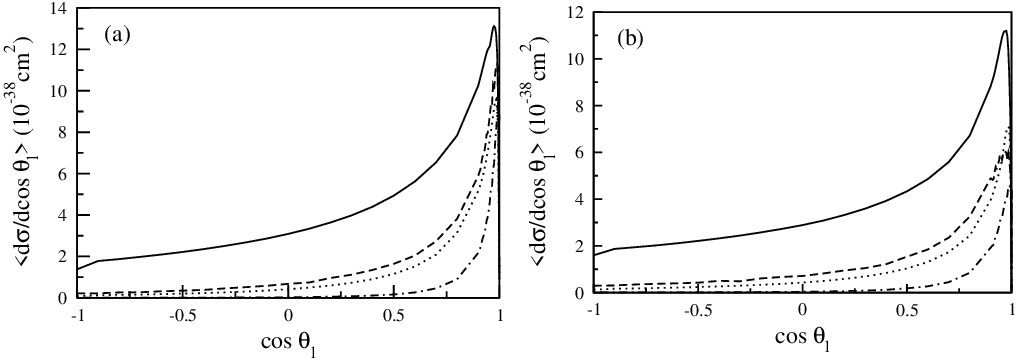}
\caption{Flux averaged differential cross section $<\frac{d\sigma}{dcos\theta_l}>$ vs $cos\theta_l$ at the Soudan site simulated by Barr et al.\cite{barr}. The solid line is for total quasielastic events and dashed line is the one $\pi$ production events for $\nu_\mu$ reaction in $^{56}{Fe}$. The corresponding results for $\bar\nu_\mu$ are shown by dotted and dashed-dotted lines, Fig.(a) for muon type and Fig.(b) for electron type neutrinos.}
\end{figure*}

\begin{figure*}
\includegraphics{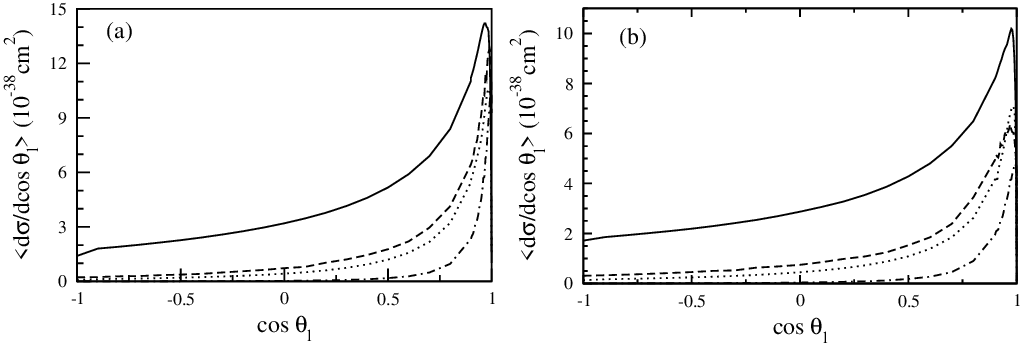}
\caption{Flux averaged differential cross section $<\frac{d\sigma}{dcos\theta_l}>$ vs $cos\theta_l$ at the Gransasso site simulated by Barr et al.\cite{barr}. The solid line is for total quasielastic events and dashed line is the one $\pi$ production events for $\nu_\mu$ reaction in $^{56}{Fe}$. The corresponding results for $\bar\nu_\mu$ are shown by dotted and dashed-dotted lines, Fig.(a) for muon type and Fig.(b) for electron type neutrinos.}
\end{figure*}

\end{document}